\documentclass[11pt,a4paper]{article}

\usepackage[utf8]{inputenc}
\usepackage[T1]{fontenc}
\usepackage{lmodern}
\usepackage[margin=2.5cm]{geometry}
\usepackage{graphicx}
\usepackage{xcolor}
\usepackage{hyperref}
\usepackage{booktabs}
\usepackage{tabularx}
\usepackage{multirow}
\usepackage{quotes}

\title{Exploring the potential of ChatGPT for feedback and evaluation in experimental physics}

\author{Marcos Abreu$^{1}$, Alvaro Suárez$^{2}$, Cecilia Stari$^{3}$, Arturo C. Marti$^{3}$\thanks{marti@fisica.edu.uy}}

\date{}

\begin{document}
\maketitle

\begin{center}
$^{1}$ Consejo de Formación en Educación, Administración Nacional de Educación Pública, Florida, Uruguay\\
$^{2}$ Consejo de Formación en Educación, Administración Nacional de Educación Pública, Montevideo, Uruguay\\
$^{3}$ Instituto de Física, Universidad de la República, Montevideo, Uruguay
\end{center}

\vspace{0.2cm}

\begin{abstract}
Recent advances in large language models have opened new possibilities for AI-assisted grading in higher education. This study explores the use of a GPT-5.4–based system to support the evaluation of introductory physics laboratory reports using a rubric-driven automated workflow implemented in batch mode via API. AI-generated scores were compared with instructor grading, and the feedback was analyzed qualitatively across rubric criteria. The results show weak agreement in the ranking of reports and noticeable differences at the level of individual scores. Item-level analysis indicates that the model frequently produced feedback classified as correct application but also generated a non-negligible proportion of reasonable but superficial responses and invalid evaluations. These limitations were mainly associated with restricted access to evidence during text extraction and OCR, particularly for equations, figures, and graphical representations. The findings suggest that batch-based AI grading can provide systematic feedback and help identify recurring patterns in student work, supporting instructors in large-scale courses. However, under the conditions examined, AI-generated scores were not interchangeable with instructor grading and should be understood as a tool to assist, rather than replace, the evaluation process.
\end{abstract}

\section{Introduction}

The emergence of Generative Artificial Intelligence (AI), particularly Large Language Models (LLMs), represents a transformative paradigm across multiple fields. Within the Physics Education Research (PER) community, Natural Language Processing (NLP) techniques were already being explored for the large-scale analysis of students’ written responses \cite{bralin2023analysis,Wilson2022Clasification}.
More recently, models such as ChatGPT have been increasingly explored for generating physics-related tasks   \cite{liang2023exploring,kieser2023educational,lopez2024challenging} and for supporting the evaluation of student work \cite{Kortemeyer2025}, showing promise in applying domain-specific assessment criteria to complex responses. Despite these advances, the pedagogically effective and responsible integration of these models within the specific domain of physics education remains a significant challenge \cite{Lademann}.

Laboratory reports are multidimensional assessment tools that integrate written explanations, mathematical reasoning, and experimental data.
They provide valuable evidence not only for assessing individual student performance but also for revealing how learners engage with authentic scientific practices—designing experiments, interpreting uncertainty, and refining models.
The evidence required for assessment may be distributed across text, equations, tables, and figures; consequently, AI-assisted evaluation may depend on what can be reliably extracted from student submissions for verification against the assessment criteria.
When students revise their reports with guided feedback, measurable growth can be observed in their scientific abilities and experimental reasoning \cite{Bugge}. However, evaluating such reports remains a demanding process, with persistent challenges regarding grading consistency and the reliability of feedback in large-enrollment physics courses \cite{passonneau2023ideal}.

While AI is reshaping many dimensions of teaching and learning \cite{zhai2023chatgpt}, its application to the evaluation of laboratory reports is still in its infancy. The complete automation of grading remains unrealistic, given the complexity inherent in validating scientific reasoning and content \cite{Kortemeyer2025}. Nonetheless, recent studies in physics education have begun to explore its potential as a complementary tool for laboratory-based assessment, including the work of Mills et al. \cite{mills2025prompting}, who examined GPT models for providing formative feedback on student reports, and Kilde-Westberg et al. \cite{kilde2025generative}, who tested generative AI as a “lab partner” in experimental activities. These efforts illustrate both the promise and the limitations of AI in supporting authentic scientific reasoning. Effective and responsible integration therefore requires maintaining a balance between automation and expert validation, not only to reduce instructors’ workload but also to preserve the integrity of academic evaluation. Within this perspective, AI opens new possibilities for tracking students’ progress over time, comparing performance across cohorts, and promoting coherence in grading among multiple instructors in large-scale physics courses.

Large language models exhibit a strong dependence on prompt design and training conditions \cite{wei2022chain}. In educational assessment, their capacity to evaluate scientific reasoning and integrity depends not only on the model architecture but also on the way humans interact with it. 
Studies in physics education have shown substantial variability across models and prompting strategies when analyzing students’ laboratory reports and explanations \cite{fussell2025comparing}. 
Developing structured prompts and standardized interaction protocols is therefore essential to achieve consistent and context-sensitive evaluations. 
A key open question is not only whether AI systems can attain a conceptually grounded understanding of students’ reasoning, but also how such depth of analysis can be effectively achieved through appropriate design and interaction strategies.

Within this framework, the present study examines the use of a large language model to support the batch assessment of physics laboratory reports. Specifically, it characterises how an AI-assisted grading system based on GPT-5.4, configured with a fixed grading rubric, performs when evaluating multiple laboratory reports, and identifies the main potentials and limitations of this approach for teaching practice. The guiding research question is: \textit{How can an AI-assisted grading system based on GPT-5.4, configured with a specific grading rubric, assist in the evaluation of laboratory reports in Experimental Physics, and what potentials and limitations emerge from its use in this context?}

To address this question, the study adopts an exploratory approach centred on a qualitative analysis of the AI-generated feedback across rubric criteria, complemented by descriptive comparisons with instructor grading to provide contextual reference.
The paper is organised as follows. First, we describe the research design and the implementation of the AI-assisted assessment protocol. Next, we present the results obtained from its application in an Experimental Physics course. Finally, the discussion and conclusions examine the strengths, limitations, and potential pedagogical implications of this exploratory experience.

\section{Research Design and Implementation}

\subsection{Framework and assessment procedure}

The study was conducted in the course Experimental Physics I at the School of Engineering of the Universidad de la República (Uruguay), which enrols approximately 300 second-year students,
As part of the course, students perform a series of experiments aimed at developing skills in measurement, data analysis, physical modelling, and the communication of results through written laboratory reports. Within this context, the aim of this work was to examine the potential of Artificial Intelligence (AI)—specifically GPT-5.4—as a tool to support the grading of laboratory reports, evaluating its ability to apply the same assessment criteria used by instructors.

The laboratory activity selected for analysis was \textit{Reaction Time and Statistics}, which aims to experimentally measure human reaction time and introduce students to the uncertainty and error analysis. The experiment is performed in pairs: one student holds a 50-cm graduated ruler vertically and releases it without warning, while the other attempts to catch it as quickly as possible after the fall begins. The measured quantity is the distance travelled by the ruler before being stopped, from which the reaction time is calculated assuming free-fall motion. Because individual measurements are affected by random variations and by the intrinsic variability of human response, the procedure is repeated several times. This enables students to analyse the distribution of results, determine a representative mean reaction time, and estimate the associated measurement uncertainty.
This activity was selected because its clearly specified guidelines tend to produce reports with a relatively uniform structure while preserving variability in reasoning and analysis. The experiment is easily replicable and involves key elements such as mathematical modelling, uncertainty propagation, statistical analysis, and graph interpretation, making it a suitable case study for evaluating AI-assisted grading.

The course uses a standardized grading rubric for each experiment (see Supp. Mat.) to ensure consistent and comparable assessment of laboratory reports. Students worked in pairs and submitted their reports as PDF files through the institutional virtual platform. To preserve realistic implementation conditions, the reports were analyzed in their original format, which included non-textual elements such as tables, equations, and graphs. All reports followed the required structure: objectives, theoretical background, experimental setup, data analysis, and conclusions.

The analysis was conducted on a random sample of 57 reports from the 2025 academic year, selected from 150 submissions. All documents were anonymized prior to processing. To ensure comparability, the reports were evaluated using the same instructions and a single batch-grading protocol implemented via the API.

\subsection{AI configuration, protocol, and analysis}

To use the AI as a grading assistant, a set of instructions was developed based on the 10-point rubric used by instructors in this laboratory activity. The rubric incorporated into the instructions corresponded exactly to the one used by the teaching staff, ensuring equivalence of evaluation criteria and facilitating comparability between human and AI assessments. The instructions defined the AI’s role as a laboratory report evaluator and specified both the assessment criteria and the expected format of the feedback. For each rubric item, the AI was required to provide the assigned score, a justification linked to the criterion, a brief statement of strengths and weaknesses, and a short overall evaluation summary. A structured checklist was also included to guide the systematic review of reports. These verifications addressed the scientific validity of the report, the correct use of equations and graphical representations, the treatment of significant figures and uncertainties, the coherence between experimental data and conclusions, and specific requirements related to the statistical analysis of the activity. In this way, the instructions sought to reproduce as closely as possible the criteria normally applied by the teaching team.

Once the grading rubric and evaluation criteria were established, they were implemented in an automated batch-grading pipeline via the API using the GPT-5.4 model. Student reports were submitted in uncompressed PDF format. For each report, text content was extracted using two complementary approaches: the PyMuPDF Python library was employed to recover the embedded digital text, while Tesseract OCR was applied to recognize characters present in graphical elements and figures. This combined extraction strategy yielded a structured textual representation of each report, which proved particularly valuable for documents containing mathematical notation and graphical content. Direct extraction of the digital text layer facilitated reliable retrieval of equations, numerical values, physical symbols, and units as they appeared in the original submissions.
As a complementary comparison, exploratory tests were also conducted by submitting the complete PDF files directly to a multimodal model. Although recent multimodal models have shown substantial improvements in visual interpretation, the exploratory trials revealed that certain complex elements—such as small figures, embedded equations, and low-resolution scanned documents—exhibited variability in their interpretation. For this reason, the text-extraction and OCR-based approach was adopted as the primary method, as it offered greater methodological stability and ensured full traceability throughout the automated analysis of student reports. The complete instruction set is provided as Supplementary Material, along with the automation script and a representative example of the system output.

The analysis comprised two complementary components. First, a quantitative comparison was conducted between the scores assigned by instructors and those generated by the AI for the 57 reports. Instructor scores correspond to the official grading performed using the standardized rubric. To evaluate the monotonic association between both grading systems without assuming linearity, the Spearman rank correlation coefficient ($\rho$) was calculated. In addition, the mean absolute error (MAE) was computed to quantify the average magnitude of discrepancies.

Second, a qualitative analysis of the AI feedback was performed to characterize how the rubric was applied and to identify sources of divergence with instructor grading. Responses were analyzed item by item and coded into three categories: correct application, when the feedback identifies the evaluated aspect and justifies the score with explicit evidence from the report; reasonable but superficial, when the feedback is plausible but does not verify specific requirements or provide traceable justification; and invalid evaluation, when the score or comment cannot be supported by the report, either because it contradicts available evidence or because relevant evidence was inaccessible or not interpretable by the model. This procedure did not involve re-grading the reports but rather classifying the AI feedback in relation to the rubric and the evidence available in the original PDF.
Coding was performed independently by two of the authors, both of them Physics teachers
with extensive experience teaching introductory experimental physics; discrepancies were resolved by consensus.

Because the reports were processed in their original PDF format and the model relied on automatic text extraction and OCR of embedded elements, an additional indicator of evidence accessibility was introduced for responses classified as invalid. This indicator distinguished between explicit limitations reported by the system (e.g., references to OCR or illegible figures) and inferred extraction problems, such as distorted mathematical notation, missing units or labels in graphs, or references to evidence present in the report but absent from the extracted text. These cases were verified by comparison with the original PDF.

Finally, a subset of cases was examined through an exploratory conversational interaction with the model, focusing on rubric items whose batch evaluation appeared limited by difficulties in accessing specific evidence (e.g., embedded equations or figures). This step was used only to better understand the system’s behaviour and did not modify the scores or coding reported in the main analysis.

\section{Results}

The AI-assisted grading protocol was applied to 57 laboratory reports from the Experimental Physics course using automated batch processing via the API. Results are presented in three stages. First, we compare the distribution of AI-generated scores with instructor grades (Subsec. 3.1). Next, we analyze the content of the AI feedback across rubric items using the analytical categories defined in the Methodology (Subsec. 3.2). Finally, we report an exploratory diagnostic reanalysis conducted in conversational mode (Subsec. 3.3).

\subsection{Comparison of scores}

Figure 1 shows the joint distribution of instructor grades and AI-generated scores for the analyzed reports (N = 57). The scatter plot exhibits substantial dispersion and no clear monotonic trend. The Spearman rank correlation coefficient ($\rho = 0.38$) indicates a weak association between both grading systems in terms of the relative ordering of reports.
On average, the AI assigned lower scores than the instructors (7.91 vs. 8.63). Differences at the level of individual reports were also observed, with a mean absolute error (MAE) of 1.01.

These indicators provide a global comparison between the two grading systems. To better understand the origin of these discrepancies, we next examine the structure and content of the AI-generated feedback across rubric items.

\begin{figure}[h]
\centering
\includegraphics[width=0.7\textwidth]{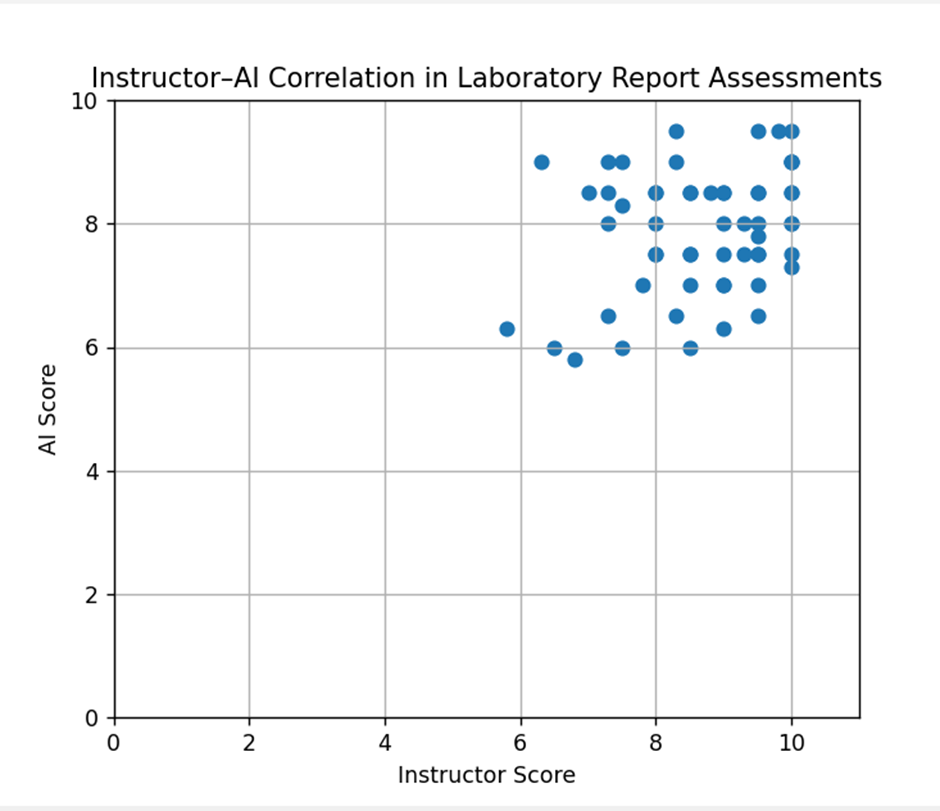}
\caption{Relationship between the scores assigned in the instructor grading and those generated by the AI system (batch grading via API). Each point represents an evaluated report ($N = 57$). Spearman’s rank correlation coefficient: $\rho =  0.38$. }\label{fig1}
\end{figure}

\subsection{Analysis of AI feedback}

AI feedback was analyzed item by item according to the rubric criteria: Objectives, Theoretical Background, Experimental Setup, Data Analysis, Conclusions, and Overall Assessment. For each item, responses were classified into three categories: correct application, reasonable but superficial, and invalid evaluation. Table 1 summarizes the percentage distribution of these categories across rubric items. Within the subset of invalid evaluations, the table also distinguishes the cross-cutting indicator of evidence accessibility into explicit and inferred limitations.

%


\begin{table}[ht]
\centering
\caption{Percentage distribution of AI-generated feedback by rubric item.
For each item, responses are classified as correct, reasonable but superficial (RS),
or invalid. Invalid responses are further disaggregated according to the type of
verification limitation: explicit  or inferred. Percentages are calculated
relative to the total number of responses for each item; therefore, in each row
the percentages of correct, RS, and invalid sum to $100\%$.}
\label{tab1}
\begin{tabular}{l c ccc|cc}
\toprule
& & \multicolumn{3}{c}{Feedback classification (\%)}
& \multicolumn{2}{c}{Evidence limitations (\%)} \\
\cmidrule(lr){3-5} \cmidrule(lr){6-7}
Item & $N$ & Correct & RS & Invalid & Explicit & Inferred \\
\midrule
Objectives & 255 & 87 & 13 & 0  & 0 & 0 \\
Theoretical background & 335 & 89 & 8 &  6 & 3 & 3 \\
Experimental setup & 360 & 84 & 10 & 6 & 3 & 4 \\
Data analysis & 820 & 84 & 5 & 10 & 4 & 6 \\
Conclusions & 252 & 85 & 5 & 11 & 2 &9 \\
Overall assessment & 739 & 81 & 12 & 7 & 4 & 3 \\
\bottomrule
\end{tabular}
\end{table}

\paragraph{Objectives}

For the Objectives item, most AI feedback was classified as \textit{correct application}. These responses typically verified whether the stated objectives were consistent with the activity and met the formal requirements of the rubric. For example, the AI noted that “the general objective of determining the reaction time is consistent with the experimental activity.” In some cases, the model also identified deviations from the task instructions, such as confusing objectives with procedural statements. A substantial fraction of responses (13\%) were classified as \textit{reasonable but superficial}. These comments generally provided broad validation without referencing specific textual evidence from the report (e.g., “the objectives are present, clear, and properly structured”), which limited the traceability of the justification.


\paragraph{Theoretical background}

For the Theoretical background item, most AI responses fell into this category. In these cases, the model recognized the relevance of the physical model presented and its connection with the experimental activity. For example, the AI noted that “the free-fall model is included with the relevant equations,” and acknowledged the presence of statistical concepts used in the later data analysis, such as Gaussian distributions and uncertainty.

Responses in this category typically consisted of general confirmations of the presence of theory (e.g., “the theory is present”) without evaluating the correctness, scope, or applicability of the model.
Invalid evaluations were mainly associated with difficulties in accessing or interpreting mathematical expressions. Both explicit and inferred limitations were observed, with inferred cases predominating. Explicit limitations included statements such as “it is not possible to verify dimensional consistency due to illegible formulas.” In several reports, the theoretical discussion referred to equations that were not recovered in the extracted text. In some instances, the AI also showed difficulty identifying certain factors in equations—for example, a  $1/2$ factor  written as a fraction rather than expressed with a horizontal division bar.

\paragraph{Experimental setup} For the Experimental setup item, most responses were classified accordingly. These comments typically referred to the level of detail in the description of the procedure and the inclusion of elements required by the rubric. For example, the AI noted that “the device and procedure are described with sufficient detail (ruler, roles, and number of measurements).” In some cases, the feedback also included suggestions aimed at improving clarity of presentation, such as indicating that “the reference position could be represented more clearly,” without implying a failure to meet the rubric criterion.

Such responses were uncommon. When present, they generally consisted of broad statements such as “sufficient description,” without specifying which aspects of the setup, materials, roles, or measurement conditions had been verified, thereby limiting the traceability of the justification.

Invalid evaluations in this item were mainly associated with explicit retrieval limitations related to figures and labels. For example, the AI reported that “the image of the setup cannot be clearly distinguished.” In several cases, the report referred to graphical evidence that was not recoverable during extraction (e.g., “As shown in Fig.~1, the sensor is placed at a fixed distance from the object”), preventing verification of the experimental configuration. Here, compared with the previous items, explicit limitations were similar  to inferred ones.

\paragraph{Data analysis} This item concentrated the largest number of AI responses (820 and  comprises three rubric subitems: (i) graphical analysis (histogram and Gaussian fit for the fall distance), (ii) determination of $y_c$ (effective fall height with its uncertainty), and (iii) determination of the reaction time $t_r$ (experimentally measured reaction time with uncertainty obtained by propagation).

Among well-grounded responses, the AI identified the presence of key analytical elements. For example, it noted that “histograms with a superimposed Gaussian are presented.” In the determination of $y_c$, appropriate metrological practices were recognized, such as reporting the value together with its uncertainty. For $t_r$, the feedback often identified the model used and the treatment of uncertainties, for instance noting that “uncertainty propagation through partial derivatives is included.” In some cases, the AI also pointed out procedural aspects affecting the traceability of the calculation, such as the absence of intermediate numerical substitutions.

Responses in the RS category typically consisted of plausible descriptions without explicit verification of the rubric criteria or traceable justification of the assigned score. For example, the AI stated that “histograms are presented and a Gaussian is mentioned,” without reconstructing the analysis or checking the specific requirements of the item.
Both explicit and inferred limitations were observed, with
slightly more inferred cases. Explicit limitations were mainly associated with the inability to reliably access graphical or tabular evidence, or labeling details required to interpret the analysis. For example, the AI noted that “it is not possible to verify whether the graphs include units,” or referred to figures mentioned in the report but not recoverable in the extracted text (e.g., “As shown in the velocity–time graph, the slope corresponds to the acceleration”). In other cases, inferred limitations were linked to distortions in the reading of equations. For instance, in the treatment of uncertainties the model failed to recognize the presence of a square root in one expression and concluded that “the type-A estimate is confused with the expression $S_n/N$,” leading to an unsupported evaluation of the statistical treatment.

\paragraph{Conclusions}

For the Conclusions item, most responses in the category evaluated the closure of the report in relation to the rubric criteria, considering its coherence with the stated objectives and reported results, as well as its integration with the preceding analysis. In these cases, the AI sometimes pointed out specific issues, for example noting that “no explicit connection with the stated objectives is observed.” In other reports, the model highlighted conclusions that linked the experimental results with uncertainty estimates and physical plausibility.

Responses in this category consisted mainly of general judgments without traceable textual evidence. Typical formulations simply acknowledged the presence of a concluding section or provided a global validation, such as “the report includes conclusions,” without specifying which elements of the text supported that assessment.

Invalid evaluations were predominantly associated with
implicit retrieval limitations. For example, the AI reported that “the coherence of the conclusion cannot be verified due to legibility issues.” In several cases, the conclusions referred to evidence not accessible in the extracted text and the model declared an inability to interpret it. For instance, a report stated: “According to the results shown in the previous graphs, the motion can be concluded to be uniformly accelerated,” but the referenced graphical evidence could not be recovered by the AI, making it impossible to verify the interpretation.

\paragraph{Overall assessment}

In this item, which summarizes a global evaluation of the reports, the percentage of responses classified as correct was the lowest among all rubric items. In line with this, responses associated with retrieval limitations were the highest. Among the responses classified as correct, the AI produced comments aligned with the rubric, for example noting that the report followed a “monograph-like structure with well-defined sections.” Responses classified as \textit{reasonable but superficial} generally consisted of global appreciations without concrete examples supporting the judgment, such as stating that the “writing is generally understandable.”

Invalid evaluations, both explicit and inferred, were also observed and were mainly associated with legibility and extraction limitations affecting multiple sections of the document. For instance, the AI reported “general legibility issues throughout the document.” In some cases, the feedback explicitly acknowledged that the overall assessment depended on evidence not recovered during automated reading, noting that the evaluation relied on the coherence between sections containing equations, figures, and graphs that were not retrieved.
Similarly, some reports included broad concluding statements—e.g., “The experiment allowed an adequate understanding of the concepts studied”—whose verification was limited by the absence of the referenced representations in the extracted text.



\subsection{Exploratory analysis in conversational mode}

As a complement to the main grading process carried out in batch mode via the API, an exploratory analysis was conducted on a subset of cases through guided interaction with the model in conversational mode. This stage was not used to modify the scores or the coding reported in the previous section. Instead, its purpose was to examine reports that, in the automated grading process, had received feedback classified as invalid evaluation, either due to explicit or inferred limitations.

In conversational mode, the interaction was organized around the specific rubric item and the corresponding evidence, requesting targeted verification when appropriate. Unlike the batch workflow, this scheme allows the analysis to focus on a single rubric item and to formulate verification requests regarding specific elements of the report, with iterative prompts when the initially retrieved evidence is insufficient or ambiguous.
In the cases reviewed, directing questions toward specific elements of the report allowed the model to consider evidence that had not been reliably incorporated in the batch workflow (e.g., embedded equations or information contained in images). As a result, the model produced feedback that no longer exhibited the verification limitations observed in the automated grading.

\section{Discussion}
We examined how an AI-assisted grading system based on GPT-5.4, configured with a rubric and implemented through automated batch processing via the API, can support the evaluation of laboratory reports and what potential and limitations emerge in this context. AI-generated scores were compared with instructor grades, and the feedback was analyzed qualitatively at the item level.
While generative AI systems have shown potential for tasks such as grading assessments and generating feedback \cite{bralin2024mapping}, our results demonstrate that their performance depends strongly on the availability and accessibility of relevant evidence in the input data.

The results show discrepancies between both grading modalities. The rank correlation ($\rho = 0.38$) indicates a weak association in the relative ordering of reports, while the mean absolute error (MAE = 1.01) reflects noticeable differences at the level of individual reports. Global averages also differ (8.63 for instructors and 7.91 for the AI), suggesting that, under the implemented conditions, AI-generated scores do not consistently align with instructor grading \cite{jukiewicz2026can,mariano2025analysis}.
Note that, although a large proportion of responses were classified as correct application at the item level, small differences accumulated across rubric criteria led to appreciable discrepancies in the final scores.

The item-level analysis helps interpret this lack of alignment. Although many responses were classified as well-grounded, the category \textit{reasonable but superficial} appeared systematically, particularly for Objectives and Theoretical background. These responses typically consisted of general validations without explicit reference to verifiable evidence in the report, even when the rubric required observable criteria. In addition, the presence of \textit{invalid evaluations} in several items introduced another source of discrepancy.
These patterns help explain how global discrepancies can coexist with locally plausible feedback. Small  differences in the application of rubric criteria—such as the degree of strictness applied to specific requirements or the traceability of the justification—may accumulate across items and become visible in global indicators such as the MAE and Spearman’s $\rho$ \cite{mariano2025analysis}.

Previous research has highlighted challenges associated with AI-assisted grading of complex student responses. For example, AI-generated scores may display overconfidence in ambiguous cases or misinterpret nuanced answers \cite{cvengros2025assisting}. In addition, tasks involving extended reasoning, mathematical derivations, or drawings remain difficult to handle reliably in automated assessment systems. These factors may contribute to the discrepancies observed between AI-generated scores and instructor grading.

A further factor concerns the automated workflow itself. API-based grading relies on text extracted from PDFs and OCR outputs of embedded content. Although large language models show potential for generating feedback in physics education, their results are not always reliable. Meyer et al. report that an open language model produced appropriate feedback in approximately 69\% of cases when evaluating physics assignments, although some responses still contained conceptual or technical inaccuracies \cite{meyer2025automatic}. Beyond these limitations, our results highlight additional challenges related to the retrieval of relevant evidence when laboratory reports are processed through OCR-based workflows.

When evaluation criteria depend on information distributed across text, equations, tables, and figures, the reliable retrieval of relevant evidence can be limited. As a result, feedback may appear plausible but insufficiently grounded in the extractable content, leading to responses classified here as \textit{reasonable but superficial} or, in some cases, to invalid evaluations. These limitations become particularly relevant when key information is contained in figures, tables, or mathematical expressions.

Within this framework, distinguishing between explicit and inferred limitations helps characterize how verification difficulties appear in the feedback. Explicit limitations correspond to declared verification impossibilities, whereas inferred limitations involve confident evaluations based on incomplete or distorted representations of the recovered content (Table~\ref{tab1}).

Finally, the exploratory conversational analysis provides additional insight into these discrepancies. In the examined cases, directing prompts toward specific rubric items and concrete elements of the report allowed the model to incorporate evidence not reliably considered in the automated pipeline, sometimes altering the justification of the feedback. This suggests that the form of the interaction is itself a relevant part of the evaluative process: in physics tasks, different prompt designs can substantially affect grading performance and, in some contexts, the quality of feedback \cite{meyer2025automatic,wei2025research}.
However, this conversational interaction was used here only as a diagnostic tool rather than as a reproducible scoring procedure, since its outcomes depend on the guidance of the interaction.

\section{Conclusion}

The results indicate that batch-based AI grading can generate systematic feedback across rubric criteria and process large sets of laboratory reports within a consistent evaluation framework. However, under the conditions examined, the scores produced by the automated workflow were not interchangeable with instructor grading. The system should therefore be understood as a tool to assist the evaluation process rather than replace instructor grading. In this sense, it may support instructors in organizing review processes, identifying recurring patterns in student work, and handling formal aspects of reports, thereby alleviating part of the routine workload associated with laboratory teaching, while keeping the instructor as the final arbiter of student performance \cite{dougru2026chatgpt}.

The main limitation concerns the traceability required in rubric-based assessment and the accessibility of evidence in the automated pipeline, which relies on automatically extracted text and OCR of embedded content. The presence of \textit{reasonable but superficial} feedback and invalid evaluations—particularly in items involving mathematical notation or graphical representations—highlights a structural constraint when the approach is used as a fully autonomous evaluation tool.






\providecommand{\newblock}{}

\end{document}